# Resolving hydrogen atoms at metal-metal hydride interfaces

*Sytze de Graaf[1,*], Jamo Momand[1], Christoph Mitterbauer[2], Sorin Lazar[2], Bart J. Kooi[1,*]*

[1]Zernike Institute for Advanced Materials, University of Groningen, Nijenborgh 4, 9747 AG Groningen, The Netherlands.

[2]Thermo Fisher Scientific, Achtseweg Noord 5, 5651 GG Eindhoven, The Netherlands.

* sytze.de.graaf@rug.nl, b.j.kooi@rug.nl

*Abstract:* Hydrogen as a fuel can be stored safely with high volumetric density in metals. It can, however, also be detrimental to metals causing embrittlement. Understanding fundamental behavior of hydrogen at atomic scale is key to improve the properties of metal-metal hydride systems. However, currently, there is no robust technique capable of visualizing hydrogen atoms. Here, we demonstrate that hydrogen atoms can be imaged unprecedentedly with integrated differential phase contrast, a recently developed technique performed in a scanning transmission electron microscope. Images of the titanium-titanium monohydride interface reveal remarkable stability of the hydride phase, originating from the interplay between compressive stress and interfacial coherence. We also uncovered, thirty years after three models were proposed, which one describes the position of the hydrogen atoms with respect to the interface. Our work enables novel research on hydrides and is extendable to all materials containing light and heavy elements, including oxides, nitrides, carbides and borides.

Hydrogen is the most abundant, but also the most light-weight element in the universe. Therefore, its direct imaging with atomic scale spatial resolution has, to date, remained elusive, despite the important role hydrogen plays in various fields such as hydrogen storage (*1, 2*) and hydrogen embrittlement (*3–5*). Nowadays, the atomic structure of most crystals can be resolved by imaging techniques based on transmission electron microscopy (TEM) and scanning TEM (STEM), that have reached resolutions well below 100 pm, after the introduction of powerful field-emission electron sources and aberration correctors (*6, 7*). However, the challenge of imaging hydrogen has remained, not only because of its low weight, but particularly due to the weight difference between hydrogen and the host atoms in the crystal. Consequently, hydrogen has not been imaged before at an interface, with vastly different concentrations of hydrogen on both sides of the interface, despite this being most interesting from a materials science perspective.

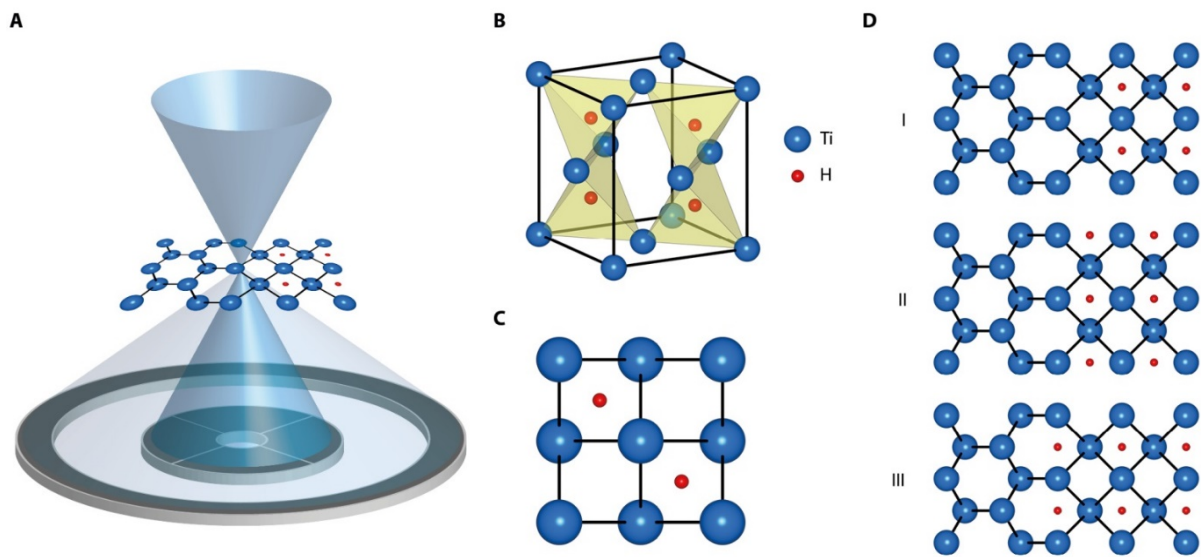

**Fig. 1: Schematic of a STEM system and the γ-TiH crystal and its three possible interfaces with α-Ti.**

*(A) iDPC images are captured with the quadrant detector (inner) and can be used simultaneously with the HAADF detector (outer). (B) Crystal structure model of the FCT γ-TiH unit cell containing four hydrogen atoms that occupy tetrahedral sites. (C) Two columns in between the titanium columns are occupied with hydrogen atoms and two columns are empty. (D) Three potential models can describe the interface between γ-TiH and α-Ti.*

In an aberration corrected STEM, the electrons are accelerated and focused into a sub-Å sized probe that scans across the surface of a thin specimen (Fig. 1A). The incident electrons interact with the specimen's local electrostatic fields as they propagate through the material and produce a scattered exit-beam of electrons that form a diffraction pattern in the detector plane. The detectors collect the electrons from a part of the diffraction pattern, of which the integrated intensity is related to the material's electrostatic field at the probe position.

The popular high-angle annular dark field (HAADF) STEM technique collects the electrons scattered to high angles using an annular detector. This technique is not affected by the wave character of the electrons that gives interferences complicating the interpretation of images. Consequently, HAADF-STEM images are readily interpretable, as atomic columns in a crystal are always imaged as bright dots in a dark surrounding with a dot brightness scaling with the average atomic number Z in the atomic column (typically $Z^{1.6-2.0}$) (*8*, *9*). Using this technique even single carbon atoms can be detected in graphene (*10–12*) and boron and nitrogen can be distinguished in 2D-BN (*13*). However, materials often consist of elements with a large difference in atomic number, such as heavy metal atoms next to light atoms like oxygen, nitrogen, carbon or hydrogen, where the scattering strength of the light elements compared to the heavy element is too low to be detected. Therefore, it is not possible to image light and heavy elements simultaneously with HAADF-STEM.

In annular bright field (ABF) STEM, the collection angles are reduced drastically in order to get sufficient signal from light elements. Nowadays, light elements such as oxygen, nitrogen and lithium can be imaged routinely with ABF-STEM (*14–16*). In the last decade this also led to the breakthrough of imaging hydrogen in bulk crystals of $YH_2$ (*17*), $VH_2$ (*18*) and $NbH_2$ (*19*). However this is only possible under special circumstances, because the specimens must be very thin, i.e. less than 10 nm, otherwise the signal from the hydrogen atoms cannot be detected. Unfortunately, not all materials can be thinned down to this level, due to limitations such as preparation induced damage layers, surface oxides or sample stability. In addition, the ABF-STEM technique automatically re-introduces the wave interference effects in the imaging, which can cause atom columns to appear white or black and generate image artefacts, making unambiguous detection of the light element challenging. Consequently, ABF-STEM cannot robustly image hydrogen atoms in realistic material systems, i.e. those that are thick and distorted, and has, therefore, been limited to model materials.

Here we image hydrogen atoms at a metal-metal hydride interface with the recently developed integrated Differential Phase Contrast (iDPC) technique. With this technique, coherently scattered electrons that fall inside the bright-field disk are collected with an annular detector that is segmented into four quadrants (Fig. 1A) (*20*). Using Differential Phase Contrast (DPC) imaging (*21–23*) three complementary images are formed that represent the material's local projected electrostatic properties when the specimen is thin. Then the projected electric field is imaged with DPC, the projected charge density with differentiated DPC (dDPC) and the projected potential with iDPC (see also supplementary materials) (*20*). The iDPC technique is particularly useful to generate sensitivity towards light elements (*24, 25*). However, the direct relation between iDPC and the material's electrostatic properties only holds in the case the specimen is ultra-thin. Usually the specimens are thicker, and the physical interpretation of the DPC images becomes less straightforward. Nevertheless, advantages of iDPC over ABF are (1) the capability to directly image the projected potential for thin specimen (2) a simpler contrast transfer function, i.e. less problems with the wave interference character and contrast reversals in the images; and (3) an intrinsically better signal-to-noise ratio (SNR), which in the end also provides the option to use lower doses e.g. on vulnerable samples.

In order to produce stable metal-metal hydride interfaces we have chosen the Ti-TiH system, that is remarkably similar to the Zr-ZrH system, which plays an important role as nuclear fuel cladding material in nuclear reactors (*26*). In contrast to earlier studies that image hydrogen in a di-hydride like $YH_2$, here we study the monohydride γ-TiH. This hydride readily forms at low temperature and low hydrogen concentrations due to the high hydrogen mobility and low room temperature hydrogen solubility limit in hexagonal close-packed (HCP) α-Ti (*27, 28*). The structure of this phase has been identified by diffraction techniques already more than three decades ago (*27–29*). The γ-TiH crystal has a face-centered tetragonal (FCT) lattice that contains four titanium and four hydrogen atoms per unit cell (Fig. 1B). Half of the tetrahedral interstitial sites are occupied by hydrogen atoms that organize in two columns that are parallel to the c-axis and are located on the face diagonal when viewed along the c-axis (Fig. 1C). Ab-initio calculations performed on the isomorphic γ-ZrH indicate that, despite a lower entropy contribution, this ordering of hydrogen in columns is more stable than the diamond-like occupation, where the hydrogen atoms occupy alternating tetrahedral sites (*30*). A similar result, although described less explicitly, was obtained for γ-TiH (*31*).

Imaging the monohydride offers major benefits over a di-hydride, because in single images we can image columns with identical surrounding of the host Ti atom columns where one type of column contains the hydrogen atoms and the other type in principle is empty. Since the wave character of the electrons plays a role during imaging it cannot be ruled out that a certain atomic-like signal is obtained in a column between Ti atom columns when no hydrogen atoms are present. In the monohydride the difference between the signals coming from the hydrogen filled columns and the empty columns can be directly compared, allowing for unambiguous interpretation that is not possible in case of the di-hydride. Imaging artefacts are often a challenge: in TEM and STEM such artefacts generally will have a symmetry related to the one of the underlying host lattices. In this respect, the monohydride offers a further advantage because the symmetry of the hydrogen sublattice is distinctly different from the one of the host titanium sublattice and therefore the weak signals we measure for the hydrogen cannot be a faint displaced or distorted replica of the host.

Even though γ-TiH has an approximately 15% larger unit cell volume compared to the α-Ti matrix, an apparently strain-free coherent interface is formed between them (*32*). Therefore, to accommodate the volume misfit, the lattice misfit perpendicular to the interface is about 16%, generating large compressive stress on the precipitate (*29*). Consequently, the growth rate parallel to the interface is orders of magnitude higher than the perpendicular one resulting in plate-shaped precipitates (*29, 32*). The absolute position of the hydrogen atoms with respect to the interface has to date not yet been identified, because diffraction techniques do not transfer translational information and real-space atomic resolution images only provided information on the position of the titanium columns (*32*). Hence, there are three possible models, schematically depicted in Fig. 1D, which could not be distinguished using earlier techniques (*32*).

To image the interface edge-on we align the α-Ti matrix along the [0001] direction and then the FCT γ-TiH is imaged along [001]. However, the large volume expansion by the incorporation of the γ-TiH precipitate in the Ti matrix leads to a severely strained system where the misfit is accommodated plastically via dislocation motion in the Ti matrix (*29, 33*). Consequently, an often encountered challenge, particularly when thinning the sample to electron transparency, is that crystal bending across the interface misaligns the crystal locally, and impedes proper atomically resolved images over the entire region of interest (Fig. S4). Due to these random imperfections in the crystal it is essential to capture high resolution images of large areas as fast as possible to locally obtain the highest possible quality atomically resolved images while minimizing drift. In this context, the detectors that we employ here have an advantage over pixelated detectors that were used in recent demonstrations of ptychographic reconstruction (*34–36*). The pixelated detector captures a complete diffraction pattern at every scan position, yielding a large 4D dataset. Then the local phase change of the electron wave can be retrieved for ultra-thin specimen, using an iterative phase retrieval algorithm. However, currently, a serious drawback of a pixelated detector is the two orders of magnitude longer pixel dwell time (*34*) compared to the HAADF, ABF and quadrant DPC detectors. Hence, despite not being a model system for imaging due to strain, bending and large sample thickness (see supplementary materials), we were able to find well aligned areas at the interface by capturing large areas with small pixel size and quick readout times of several seconds. With this approach we routinely imaged the edge-on interface with minimal residual misalignment with about 60 pm resolution (Fig. 2).

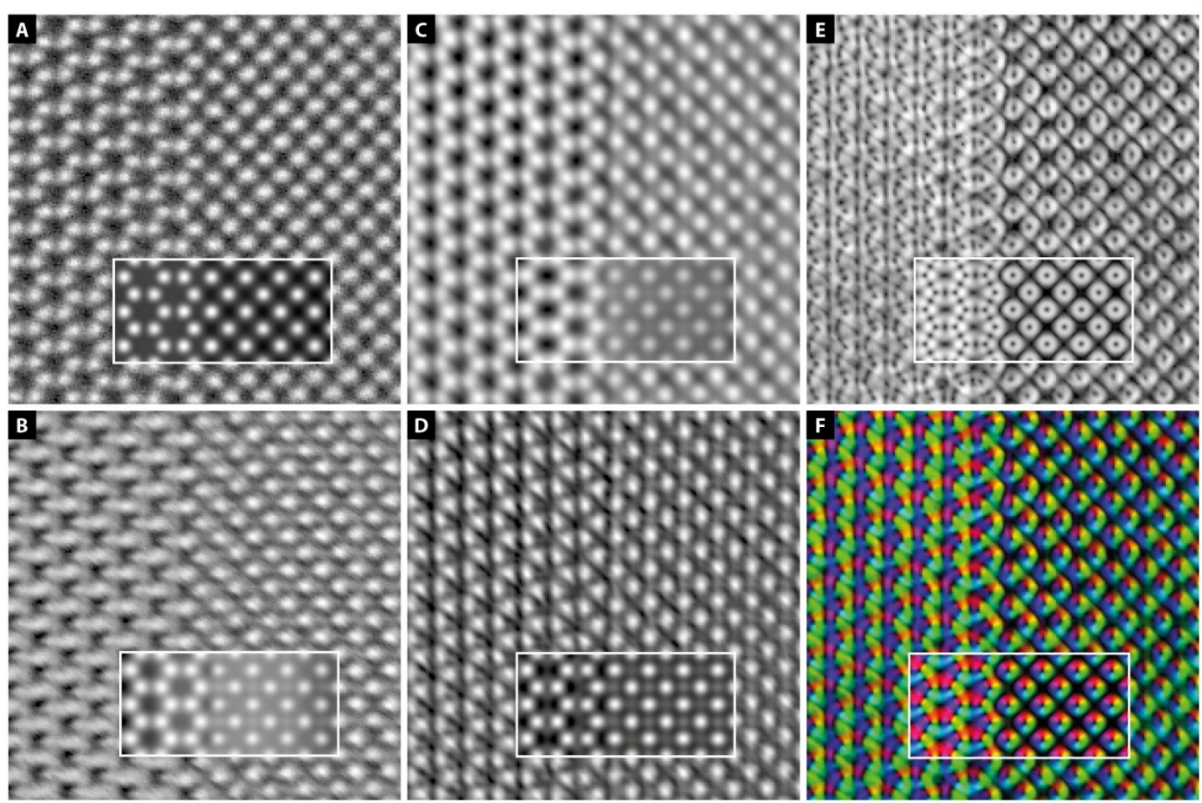

**Fig. 2: Comparison of images of the interface between γ-TiH and α-Ti using different techniques.**

*(A) HAADF. (B) Contrast inverted ABF. (C) iDPC. (D) Contrast inverted dDPC. (E) DPC magnitude. (F) DPC vector field using color wheel representation. Simulated images of a 30 nm thick specimen are inset in the image. Field of view is 3.5x3.5 nm.*

The image recorded with the HAADF detector can be interpreted directly by virtue of Z-contrast: it shows two clearly distinguishable lattices separated by an atomically sharp interface that matches convincingly with the simulated image (Fig. 2A). Viewed along these crystal axes, the α-Ti(01-10) planes are parallel to the γ-TiH(1-10) planes at the interface, in agreement with the TEM studies that were performed over three decades ago (*29*, *32*). Although the Z-contrast image visualizes the titanium atoms well, it fails to transfer signal from the hydrogen atoms, which are expected in equivalent stoichiometry in the γ-TiH.

Next we used ABF to image the α-Ti/γ-TiH interface (Fig. 2B), although the specimen thickness of about 30 nm is substantially thicker than 10 nm, which is required to properly image hydrogen atoms with ABF. The titanium atoms are well-resolved in the γ-TiH but poorly in the α-Ti. The high sensitivity of ABF imaging towards crystal orientation (see Fig. S11) combined with the inherent local bending of the specimen leads to the lower quality of the ABF image compared to the HAADF image (*15*). Nonetheless, in the γ-TiH a weak contrast is visible between alternating atomic planes parallel and perpendicular to the interface that indicates the presence of signal from the hydrogen columns. Close inspection suggests that either model I or III best describes the interface (Fig. 1D). However, being limited by the SNR and quality of the image we cannot reliably determine the exact position of hydrogen columns.

As a final step, the DPC-based images of the interface were constructed using the segmented detector and are depicted in Fig. 2C-F. In the iDPC image the Ti atoms are resolved accurately in the γ-TiH as well as the α-Ti matrix (Fig. 2C). Besides the bright Ti atoms, there is also a clear signal within the γ-TiH crystal that forms a checkerboard-like pattern with the symmetry that is expected for the hydrogen sublattice. With the convincing match between the experimental and simulated image, and the extensive additional checks that we have performed (see supplementary materials e.g. providing different types of electron energy loss spectroscopy information and results of extensive image simulations), we show that the hydrogen columns are indeed imaged with iDPC and that the interface is best described by model I (Fig. 1D).

In the dDPC image (Fig. 2D) the Ti atoms are properly resolved throughout the entire image, in contrast to the ABF image where they were only well resolved in the γ-TiH. Also the hydrogen atom columns are imaged, but, as in the ABF image, the faint signal is preventing accurate hydrogen column detection in the image. The DPC images in Figs. 2E and 2F show, respectively, the scalar field magnitude and the vector field, where the field direction is indicated by the color and the magnitude by the intensity. Fig. 2E displays a rather complex field in the Ti matrix, and a more easily interpretable field in the γ-TiH which agrees well with the simulation. The field magnitude at the Ti atom columns in the γ-TiH is not radially symmetric, but ellipse shaped and oriented in two orthogonal directions. This is a direct effect of the checkerboard-like ordered hydrogen columns and also allows us to confirm model I as the best description of the interface.

We have performed a SNR analysis of the hydrogen signal as a measure to quantify the hydrogen imaging capabilities of the different imaging techniques (Fig. 3). Note that we have applied an identical filtering procedure to all experimental images to ensure a fair comparison (see Fig. S6). Experimental intensity profiles are extracted by averaging the alternating (220) atomic planes (perpendicular to the interface) containing hydrogen columns (Ti-H-Ti), and the intermediate planes containing empty columns (Ti-empty-Ti). These are compared with the simulated intensity profiles of a 32.2 nm thick specimen (see Fig. 3), but note that the iDPC images compare reasonably well with simulations holding for a large thickness range of 20-50 nm (Fig. S13). We define the SNR as the ratio of the relative intensity of the hydrogen signal and the standard deviation of the averaged intensity profiles.

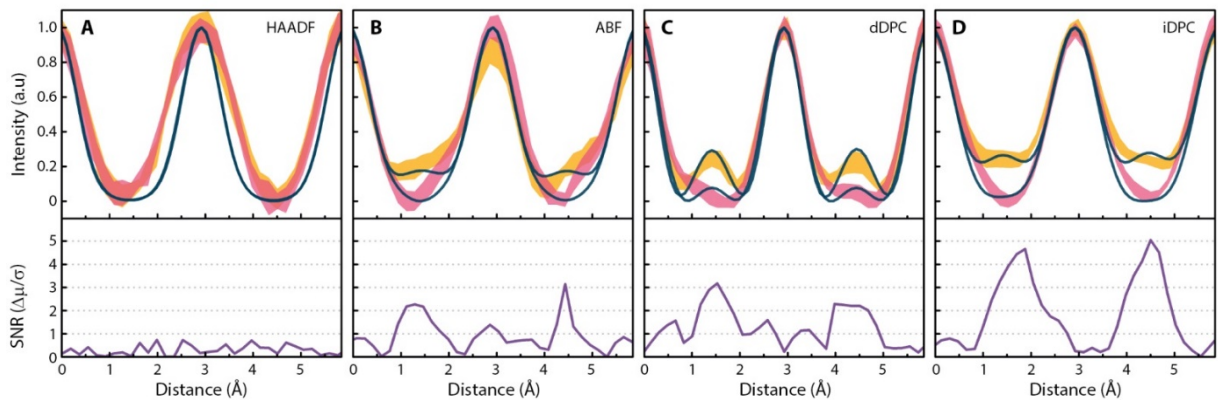

**Fig. 3: Comparison of the experimental and simulated intensity profiles of the γ-TiH at the metal-metal hydride interface, and the experimental signal-to-noise ratio of the hydrogen column signal.**

*(A) HAADF. (B) ABF. (C) dDPC. (D) iDPC. The pink and yellow bands represent the experimental intensity profiles of the Ti-empty-Ti-empty-Ti and Ti-H-Ti-H-Ti columns, respectively. The width of the bands is twice the standard deviation, centered around the average value. Solid dark lines are the simulated profiles for a 32.2 nm thick γ-TiH crystal. The SNR of the hydrogen signal is extracted from the experimental profiles and plotted below.*

The HAADF intensity profiles only contain peaks from Ti and no signal from hydrogen, hence, as expected, the SNR is below unity (Fig. 3A). In the ABF image the hydrogen atom columns are imaged with a SNR of 2-3 (Fig. 3B). However, also an image artefact is revealed: the intensity of adjacent Ti atom columns are modulated with the symmetry of the checkerboard-like ordered hydrogen atoms. The SNR is also 2-3 in the dDPC image, where the hydrogen atoms in fact produce the sharpest peak (Fig. 3C). However, at the same time, peaks are present in the empty columns that complicate image interpretation. Finally, the intensity profiles from the iDPC image show that hydrogen atoms are imaged with the highest SNR of 4-5, and no image artefacts are present (Fig. 3D). The SNR of iDPC compared to ABF is improved, because on the one hand the segmented detector collects more electrons, while on the other hand the random noise is suppressed by its intrinsic noise suppression property (*20*, *37*). Furthermore, the inherently present specimen distortion reduces signal from the hydrogen columns for ABF and substantially less for iDPC (see Fig. S7 and S11).

We did not observe any structural rearrangements of hydrogen atoms at the interface during the imaging, i.e. we were limited by carbon contamination and not by electron beam induced damage. This is a remarkable observation, considering that (i) energetic electrons are impinging on the specimen during imaging, that can either transfer energy to the material, or displace hydrogen or titanium atoms upon knock-on collisions (*38*), (ii) hydrogen atoms are highly mobile in titanium at room temperature, (iii) there is a step function in hydrogen concentration of about 50 atomic percent across the interface and (iv) the γ-TiH is a metastable phase. Therefore, our observation is at first sight contrary to what is expected based on these considerations. However, energy transferred by inelastic collisions of the 300 keV primary electrons with the atoms can be quickly dissipated because of the metallic behavior of titanium hydrides (*39*). Knock-on collision of a primary electron with a hydrogen or titanium atom can lead to a maximum energy transfer of 844 eV and 17.8 eV, respectively. This amount of energy is sufficiently high to displace hydrogen atoms but not titanium atoms, if we assume typical binding energies of 25 eV (*40*). Nevertheless, overall, the probability to displace a hydrogen atom may still be low, because the scattering cross section of the hydrogen atom is small and it decreases with increasing primary electron energy (*38*). Therefore, it is advantageous to image the titanium hydride with higher electron energies (like 300 keV) to minimize electron beam induced damage. Furthermore, for a complete understanding it is also critical to consider the effects of embedding the γ-TiH precipitate in the host metal. Ab-initio simulations (*31*) have demonstrated that the γ-TiH precipitate cannot be stabilized by compressive stress imposed by the matrix alone. Only when the combined effect of compressive stress imposed by the matrix, and the coherency strains of the interface, are taken into account, the γ-TiH precipitate is stabilized. These arguments rationalize our observation that the Ti/γ-TiH interface remains stable during imaging and electron beam induced damage is not visible. We additionally observed that after preparation, the γ-TiH remains only crystalline in reasonably thick (e.g. 30 nm or thicker) parts of the specimen. We speculate that the thinner parts of the specimen are easier amorphized during preparation, as it may not have the mechanical rigidity to provide sufficient coherency strains and compressive stresses that stabilize the γ-TiH. Hence, the dynamic interplay between compressive stress and the formation of a coherent interface is directly responsible for the high stability of the Ti/γ-TiH interface.

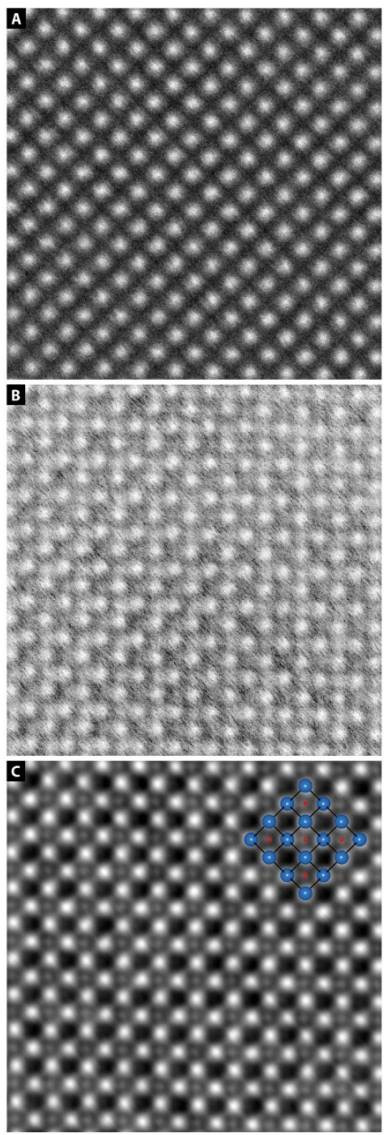

**Fig. 4: Comparison of high-quality images of γ-TiH far away from the interface.**
*(A) HAADF. (B) Contrast inverted ABF. (C) iDPC. Field of view is 3.13x3.13 nm.*

Higher quality images of the γ-TiH are captured further away from the interface where the crystal strain has relieved and local bending is much less a problem (Fig. 4). A result is that the hydrogen columns appear more localized compared to the images that are recorded at the interface. Now the checkerboard-like filling of the hydrogen columns in the γ-TiH lattice is visible in the ABF (Fig. 4B), and iDPC (Fig. 4C) image which is lacking in the HAADF-STEM image (Fig. 4A). Here we constructed the ABF image by summing the four images from the quadrant detector, because with the normal ABF geometry a signal from the hydrogen atoms could not be detected. A similar detector geometry was also used to image hydrogen in $VH_2$ (*18*). Our simulations show that in this case a higher hydrogen column intensity can be achieved at the costs of localization (Fig. S7). Indeed, now the hydrogen columns are also readily observable next to the titanium ones in the ABF image. Nevertheless, the hydrogen atoms are much better visualized in the iDPC image.

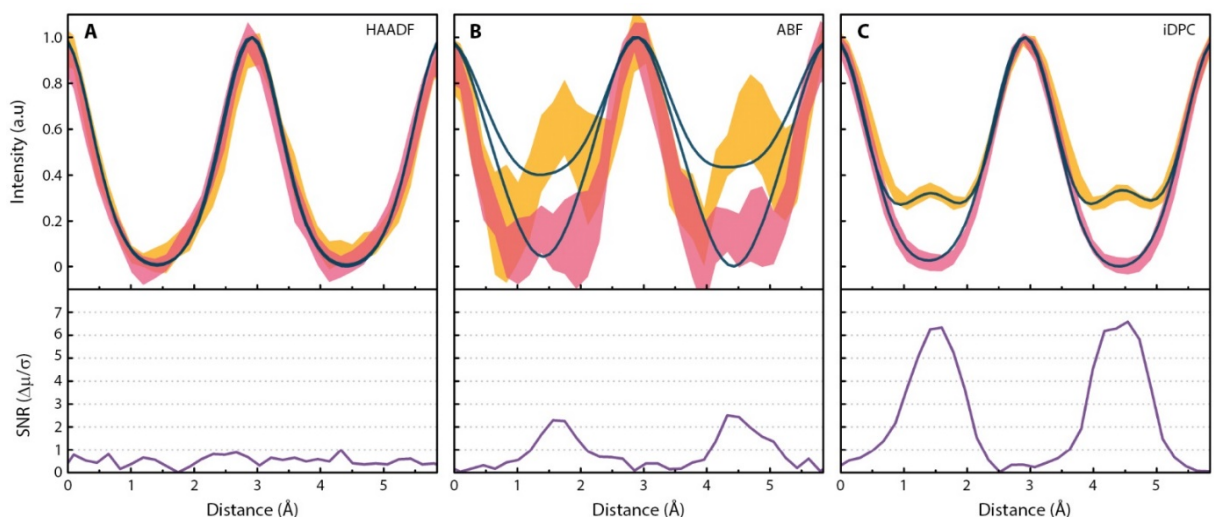

**Fig. 5: Comparison of the experimental and simulated intensity profiles of the γ-TiH under optimal imaging conditions, and the signal-to-noise ratio of the hydrogen column signal.**

*(A) HAADF. (B) ABF. (C) iDPC. Here the ABF is formed by summing the quadrants of the iDPC detector, such that the ABF and iDPC image use the exact identical raw data. The pink and yellow bands represent the experimental intensity profiles of the Ti-empty-Ti-empty-Ti and Ti-H-Ti-H-Ti columns, respectively. The width of the bands is twice the standard deviation, centered around the average value. Solid dark lines are the simulated profiles for a 52 nm thick γ-TiH crystal. The SNR of the hydrogen signal is extracted from the experimental profiles and plotted below.*

Compared to Fig. 3, the experimental intensity profiles extracted from the images in Fig. 4 now match significantly better with the simulated ones as the crystal is better oriented (see Fig. 5). Particularly the profiles extracted from the HAADF and iDPC images agree nearly perfectly with the simulations. However, the intensity profile from the ABF image does not match well and contains high levels of noise. In fact, the noise in the iDPC image is three times lower than in the ABF image, which is a remarkable result, considering that the ABF and iDPC images are constructed from the exact identical original signal. This demonstrates the intrinsic noise suppressing property of using in a clever way the four quadrants of the iDPC imaging technique.

Thus, our experimental images and extensive image simulations of the γ-TiH unit cell (see supplementary materials) allow us to conclude that iDPC has several major advantageous over ABF imaging: (i) iDPC images the hydrogen columns with higher contrast and better localization. (ii) The atom column intensity oscillates substantially less as a function of specimen thickness. (iii) Contrast also does not reverse as long as the probe is focused into the specimen, unlike ABF imaging where the empty column and hydrogen filled column reverse contrast for specimen thinner than about 10 nm. This furthermore shows that the iDPC technique produces easily interpretable images also when the specimen is thick. However, then the iDPC image cannot be quantitatively interpreted as the material's projected potential. This is also evident, since we observe in experiment and in simulation that the atom intensities oscillate as a function of thickness, and also that the relative contrast of the hydrogen atoms is higher than expected based on the atomic weights of hydrogen and titanium. Nevertheless, these simulation results clearly demonstrate that on fairly all aspects iDPC images allow more straightforward interpretation and better imaging of the light element column next to an empty column. In ABF, the signal coming from the hydrogen column and the empty column are very similar and can reverse in contrast which shows that ABF is prone to erroneous detection of light elements.

In summary, we have imaged hydrogen atoms in titanium monohydride at its interface with titanium using the recently developed iDPC-STEM technique. The images of the interface uncovered, thirty years after three models were proposed, the model that describes the positions of the hydrogen atoms with respect to the interface. We observed unexpected high stability of the titanium monohydride even during energetic electron irradiation conditions, originating from the combined effect of interfacial coherency strains and the compressive stress imposed by the matrix. Significantly, the capabilities of imaging hydrogen in the non-ideal titanium-titanium monohydride system, with severe and limiting imperfections, demonstrate the ability and the prospects of iDPC-STEM as a robust imaging tool in materials research. This work paves the way to further advance our understanding of hydrogen in solids, like retrieving local site occupancy, atomic vibrations and mobility from iDPC images. Moreover, it can be extended to all materials systems containing light elements next to heavy ones, like oxides, nitrides, carbides and borides.

## Methods

### Specimen preparation

A single crystal titanium sample was mechanically polished followed by twin-jet electrochemical polishing at room temperature with a TenuPol-3 from Struers, to load the sample with hydrogen such that the γ-TiH precipitates were formed. A TEM lamella was extracted from the sample using a Helios G4 CX dual beam system with a Ga focused ion beam. The lamella was thinned to electron transparency with the focused ion beam using progressively lower accelerating voltages. As a final step a Gatan PIPS II polishing system was used to polish the lamella with 0.3 kV Ar ions.

### Scanning transmission electron microscopy

For the imaging we used a probe and image corrected Thermo Fisher Scientific™ Themis Z S/TEM system operating at 300 kV. The specimen was plasma cleaned for three minutes prior to insertion in the S/TEM column. For the imaging a convergence semi-angle of 21 mrad was used, and the current was set to 50 pA for Fig. 2 and 14 pA for Fig. 5. Experimental images are identically filtered by applying a high-pass Gaussian filter and an average background subtraction filter.

### Multislice simulations

The crystallographic models of the γ-TiH unit cell and the γ-TiH/α-Ti interface were constructed with VESTA. The γ-TiH unit cell was adjusted (< 2% change in lattice parameters) to obtain a coherent interface with α-Ti. These models were loaded in the Dr.Probe software for STEM image simulations. Microscope parameters were set equal to experimentally calibrated values, and aberrations were neglected except for defocus. The detectors collection angles were also set to the experimentally calibrated values. The resulting images were convolved with a two-dimensional Gaussian function of 70 pm FWHM to account for finite probe size.

**Acknowledgments**

Financial support from the Zernike Institute for Advanced Materials and the Groningen Cognitive Systems and Materials Center is gratefully acknowledged. In particular prof. B. Noheda and prof. M.A. Loi are acknowledged for supporting the TEM facility. G.H. ten Brink is acknowledged for technical support.

Supplementary materials for:

# Resolving hydrogen atoms at metal-metal hydride interfaces

*Sytze de Graaf[1,*], Jamo Momand[1], Christoph Mitterbauer[2], Sorin Lazar[2], Bart J. Kooi[1,*]*

[1]Zernike Institute for Advanced Materials, University of Groningen, Nijenborgh 4, 9747 AG Groningen, The Netherlands.

[2]Thermo Fisher Scientific, Achtseweg Noord 5, 5651 GG Eindhoven, The Netherlands.

* sytze.de.graaf@rug.nl, b.j.kooi@rug.nl

## 1. Specimen preparation and characteristics

Disks with a diameter of 3.05 mm diameter and a thickness of about 350 μm thick were spark eroded from a titanium single crystal having a purity of 99.99% (as purchased from Goodfellow). The disks were mechanically polished on both sides to a final thickness of about 50-60 μm using sequentially 1000, 2000 and 4000 grit SiC grinding paper. The mechanically polished disks were twin-jet electrochemically polished at room temperature using a TenuPol-3 from Struers. For the electrolyte a mixture of 600 ml methanol, 360 ml 2-butoxyethanol and 64 ml of perchloric acid (72%) was used. The applied bias voltage was 10 V, with maximum flowrate (10) of the electrolyte and maximum photosensitivity (10) of the photodiode. This resulted in a constant polishing current of 100 mA for 20 seconds until the polishing device detected perforation and stopped the polishing.

During this process of electrochemical polishing plate-shaped precipitates are produced within the titanium which (in a later stage) can be identified as γ-TiH. This process of hydride formation during electrochemical polishing is well-known (*29*, *32*, *41*, *42*). During this process the surface oxide on the titanium is continuously etched away into the liquid allowing hydrogen dissolution in the solid titanium. Since the solubility limit of hydrogen in Ti is low, the hydrogen easily precipitates in the form of thin γ-TiH platelets. We spent considerable time in characterizing and optimizing the electrochemical polishing process in order to obtain a desired structure with relatively long planar interfaces of the γ-TiH platelets in the Ti matrix without too much stresses and bending of the thin foil.

We analyzed the crystal orientation of the disk using electron backscatter diffraction (EBSD) in an FEI Nova NanoSEM 650 equipped with an Ametek EDAX-TSL EBSD system. A result from this analysis is shown in **Figure S1**; it can be concluded that the sample consists mostly out of a millimeter sized grain with the c-axis tilted about 35° from the plane normal. However, there are also micrometer thick bands, extending throughout the millimeter sized crystal, that have the c-axis approximately in plane.

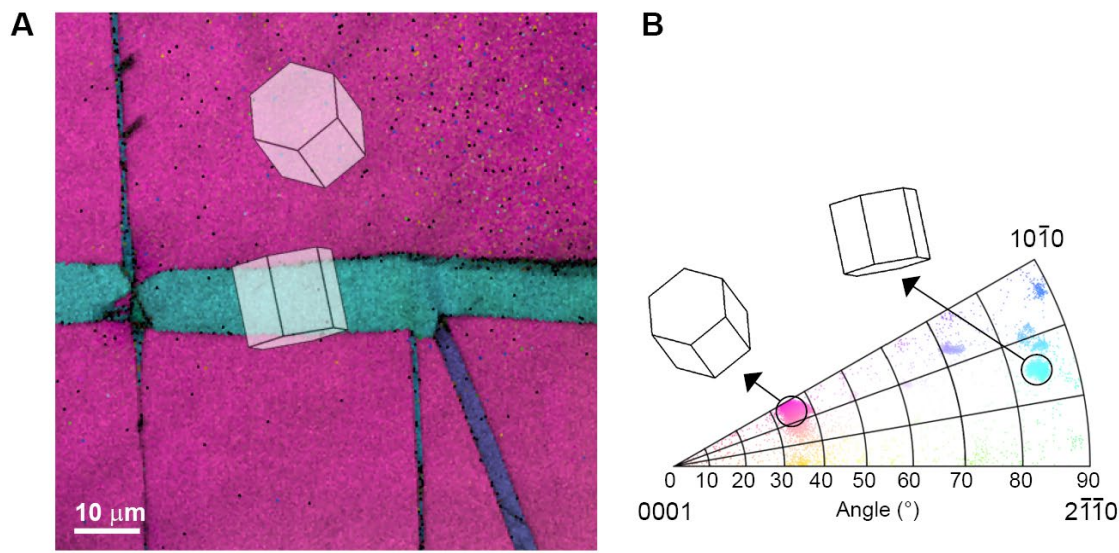

*Fig. S1. Electron back-scatter diffraction (EBSD) map of the Ti sample. (A) Orientation imaging microscopy image reveals two dominant grain orientations. The corresponding inverse pole figure (B) shows that the c-axis of the larger (purple) grain is tilted about 35° with respect to the plane normal and the thin (turquoise) band has its c-axis 85° from the plane normal, or 5° from being in plane.*

To obtain specimen with the desired c-axis normal to the sample plane we extracted lamellae from these thinner bands using a Helios G4 CX dual beam system with a Ga focused ion beam (FIB). A lamella was attached to a Cu TEM grid and thinned to electron transparency using progressively lower accelerating voltage from 30 kV down to 1 kV. During this stage the plate-shaped precipitates can be observed readily and they even show some further growth. The FIB thinned lamella was finally polished for 60 seconds on each side using a Gatan PIPS II Argon ion polishing system operating in stationary mode at 0.3 kV with gun angles of ±10°. Prior to S/TEM imaging, we plasma cleaned the TEM grid with the FIB lamella for three minutes with a Fischione Model 1070 plasma cleaner, using a gas mixture of 25% O2 and 75% Ar.

An overview image of the γ-TiH precipitates is shown in **Figure S2A**. Despite the careful thinning procedure, we observed with S/TEM that the γ-TiH precipitates are amorphous close to the edge of the lamella where the foil is the thinnest (see **Figure S2B**). As a consequence we were forced to image the interface relatively far away from the edge. For example, the images of the interface that we present in Figure 2 of the main article are captured from an area indicated by the white rectangle in Fig, S2A. This area is approximately 100 nm away from the edge, and we estimate the local thickness to be in the range of 30-40 nm based on careful comparison with image simulations.

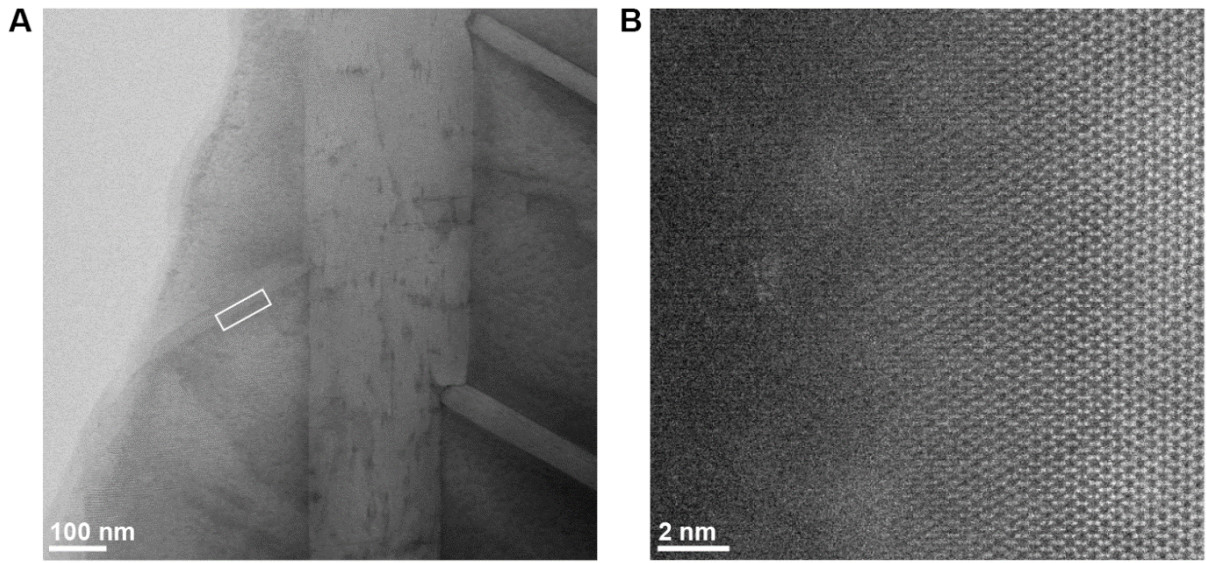

*Fig. S2. Amorphous g-TiH in thin parts of the specimen. (A) Overview ABF-STEM image shows the γ-TiH precipitates in the α-Ti matrix. The images of the interface (Fig. 2 in the main article) were captured in the area indicated by the white rectangle. (B) The HAADF-STEM image of the interface of the central thickest precipitate that was captured close to the thin edge of the lamella. It shows that the γ-TiH precipitate (on the left) is amorphous unlike the α-Ti host (on the right) which remains crystalline.*

In bright-field TEM analysis we observe bending contours that extend far into the specimen (i.e. covering all areas that are still relatively thin). The crystal bending is a feature of the specimen that is inherent to the incorporation of the γ-TiH precipitates, with about 15% larger unit cell volume compared to the host Ti. With the Kikuchi patterns that we used to align the crystal we observed that the smallest length scale of the crystal bending is of the order of several nanometers. The crystal was also bent across the lamella with a periodicity of the order of hundreds of nanometers; however this is irrelevant considering the used field-of-view during imaging. After aligning the crystal as good as the specimen allowed at low magnification (with a field-of-view several hundreds of nanometers), we used the Kikuchi pattern to find areas close to the interface that were correctly aligned. With this approach we were able to capture the interface with atomic resolution; however a new issue that we encountered many times is that only one of the two phases, either the α-Ti or the γ-TiH, was properly resolved.

An example that shows the crystal misalignment along the interface is depicted in Figure S3. Here the interface was imaged with atomic resolution using HAADF-STEM **(Figure S3A)** and iDPC-STEM (**Figure S3B**) simultaneously. In the HAADF-STEM image both crystals are resolved with high quality at the top of the image, but the quality starts to degrade from around the center towards the bottom of the image. We note that the interface is not atomically sharp; however this is not necessary to demonstrate the point here. At the lower part of the image the closest Ti columns in the host matrix are not separately resolved, but form streaks perpendicular to the interface. This is an effect of local stresses in the system (especially perpendicular to the interface where there is a 16% lattice mismatch between the precipitate and the matrix) that induce crystal bending such that the close Ti columns form streaks in projection. The misalignment of the Ti matrix is more apparent in the iDPC-STEM image, where only in the top 5 nm the $\alpha$-Ti and $\gamma$-TiH are of rather good quality. In the remaining part of the image the Ti matrix forms streaks similarly to the HAADF-STEM image.

Considering all these characteristic imperfections of the specimen, we had to find areas where the local stresses were minimal such that both phases at the interface could be resolved with atomic resolution. In practice this means that we (1) aligned the crystal locally at relatively low magnification as good as possible (2) captured several images with high resolution (e.g. 1024x1024 pixels up to 4096x4096 pixels) and step sizes of 5 up to 25 pm with a pixel dwell time of 5-10 µs (3) selected well aligned parts of the image based on the combined quality of the simultaneously acquired HAADF and iDPC or ABF image.

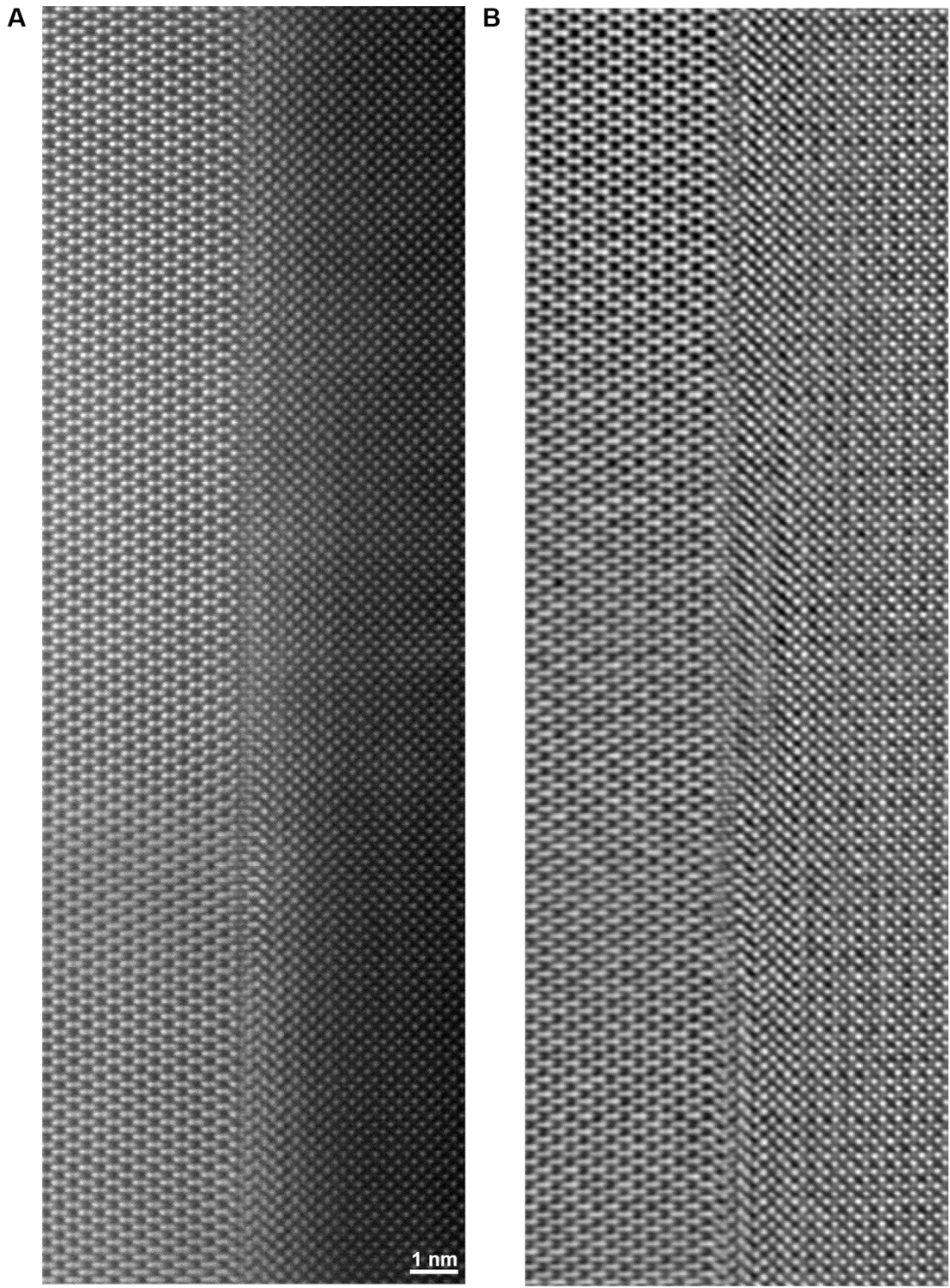

***Fig. S3. Crystal bending along the interface.*** *(**A**) HAADF-STEM image shows that the Ti matrix gradually misaligns from top to bottom (which is about 25 nm). (**B**) In the iDPC-STEM image only the top 5 nm is well-aligned. Note, however, that the hydrogen columns are properly resolved in the γ-TiH precipitate along the entire interface.*

## 2. Identification of the γ-TiH phase with electron energy loss spectroscopy

We acquired spatially resolved electron energy loss spectra (EELS) of areas containing the Ti matrix and the γ-TiH precipitates with a Thermo Fisher Scientific™ Themis Z S/TEM system operating at 300 kV equipped with a Gatan Enfinium 977 system. Spectra were extracted from equally sized areas in the Ti matrix and γ-TiH precipitate (**Figure S4**). From the low electron energy loss part we obtain the locations of the plasmon peaks in Ti and in γ-TiH at 17.6 eV and 19.4 eV, respectively, which matches well with previous studies (typically 17.6 eV and 19.6 eV) (*42, 43*) and thus can be used as a verification that γ-TiH precipitates are formed.

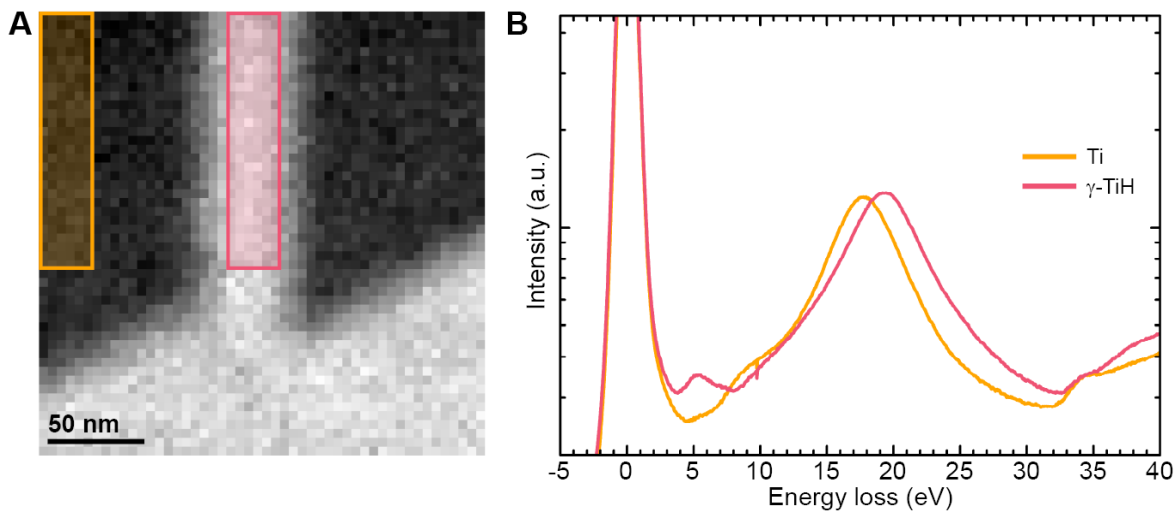

*Fig. S4. Plasmon mapping of Ti and γ-TiH.* (A) *Intensity map (with an energy window from 19.80 eV to 19.85 eV) shows the area from which we extracted the electron energy loss spectra. The γ-TiH precipitates appear with higher intensity compared to the darker Ti matrix due to the filtering.* (B) *The equally sized yellow and pink areas in (A) are integrated to yield the spectra for respectively α-Ti and γ-TiH. The peaks close to 20 eV correspond to the plasmon peaks, and are a fingerprint for the two phases.*

From the high electron energy loss part of the spectrum we investigated if contaminants/dopants are present because titanium has a high affinity for carbon, nitrogen and oxygen (**Figure S5**). The spectrum only contains an ionization edge starting at 456 eV which belongs to Ti $L_{2,3}$ edge (462 eV and 456 eV, respectively), and no signs of carbon, nitrogen and/or oxygen are present in the spectrum, because the C K edge (284 eV), N K edge (401 eV) and O K edge (532 eV) are missing. Hence, based on the EELS analysis we conclude that titanium hydride precipitates are present without any noticeable contaminants/ dopants.

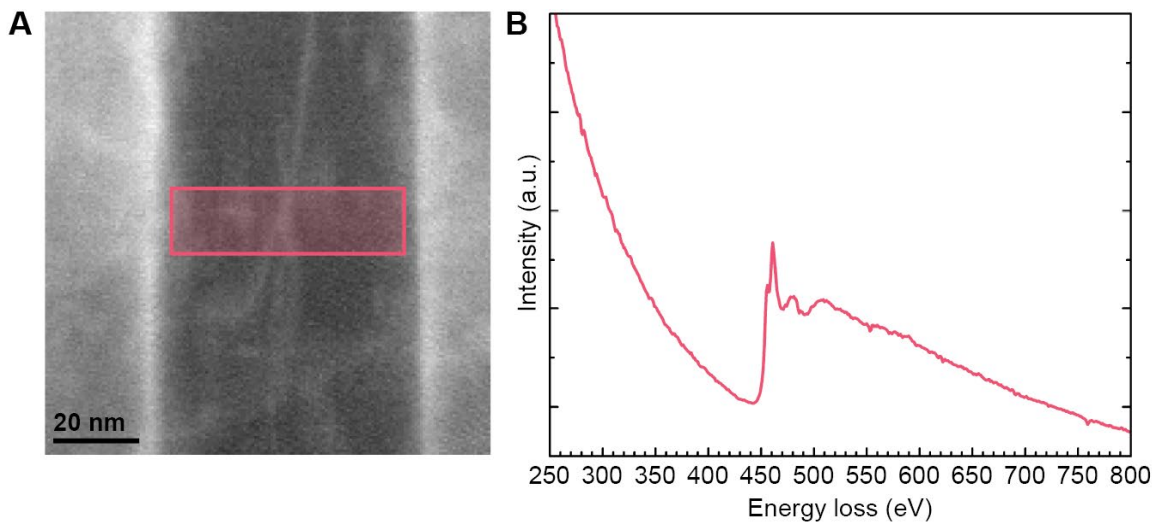

*Fig. S5. Inspection of contamination with core loss EELS. (A) Overview image of the area from which we acquired an electron energy loss spectrum. The actually scanned area of the spatially resolved spectrum was the same height as the pink highlighted area, but slightly wider to extend into the Ti matrix. (B) Integrated high-loss spectrum from the pink area in the γ-TiH precipitate as depicted in (A) is plotted. The absorption edges of C, N, O and Ti are located in this energy range but not detected as only the sharp edge from Ti is visible.*

### 3. Scanning transmission electron microscopy

For the atomic resolution imaging we used a probe and image corrected Themis Z S/TEM operating at 300 kV. Images were recorded with convergence semi-angle of 21 mrad and a probe current of 50 pA (Figure 2 of the main article) and 14 pA (Figure 4 of the main article), a step size between 5 and 20 pm, image resolution of 1024x1024 pixels up to 4096x4096 pixels and a pixel dwell time of 1 up to 10 µs.

### 4. Filtering procedure of the experimental images

All experimental images have been filtered identically with a high-pass Gaussian filter and an average background subtraction filter (ABSF) (*44*) (freely available at http://www.dmscripting.com/hrtem_filter.html). The high-pass Gaussian filter (with a standard deviation of 1.1 nm$^{-1}$, equivalent to 2.1 mrad) only attenuates information at low spatial frequencies and retains high frequency information e.g. this filter removes visualized carbon contamination and thickness variations, but keeps the atomic structure untouched. The average background subtraction filter removes background information (like noise and diffuse amorphous rings) from the entire Fourier spectrum of the image. The effect of the filters on the HAADF, ABF and iDPC images is shown in **Fig. S6**.

The high-pass Gaussian filter is particularly helpful for the techniques that transfer low spatial frequency information i.e. HAADF and iDPC. However, this filter has most effect on the iDPC image, since carbon contamination is best visualized in the iDPC image as it is more sensitive to light elements than HAADF. It has a minor effect on the ABF image, as it does not image low spatial frequency information. The ABSF enhances the contrast of all images, but has most effect on the HAADF and ABF images. The high-contrast raw HAADF image is already interpretable, and this is further enhanced by noise removal due to the ABSF. The ABSF is, however, crucial for the ABF images, because these images are noisy, which obscures the crystalline structure to a large extend. The iDPC image is improved as well, however the carbon contamination remains.

For a fair comparison between the images it is therefore required to combine the two filters. In that case the HAADF and ABF are mainly enhanced by the ABSF, and the iDPC image by the high-pass Gaussian filter. This also gives the best average signal-to-noise ratio (SNR) of the hydrogen signal in the ABF and iDPC images (see insets in Fig. S6).

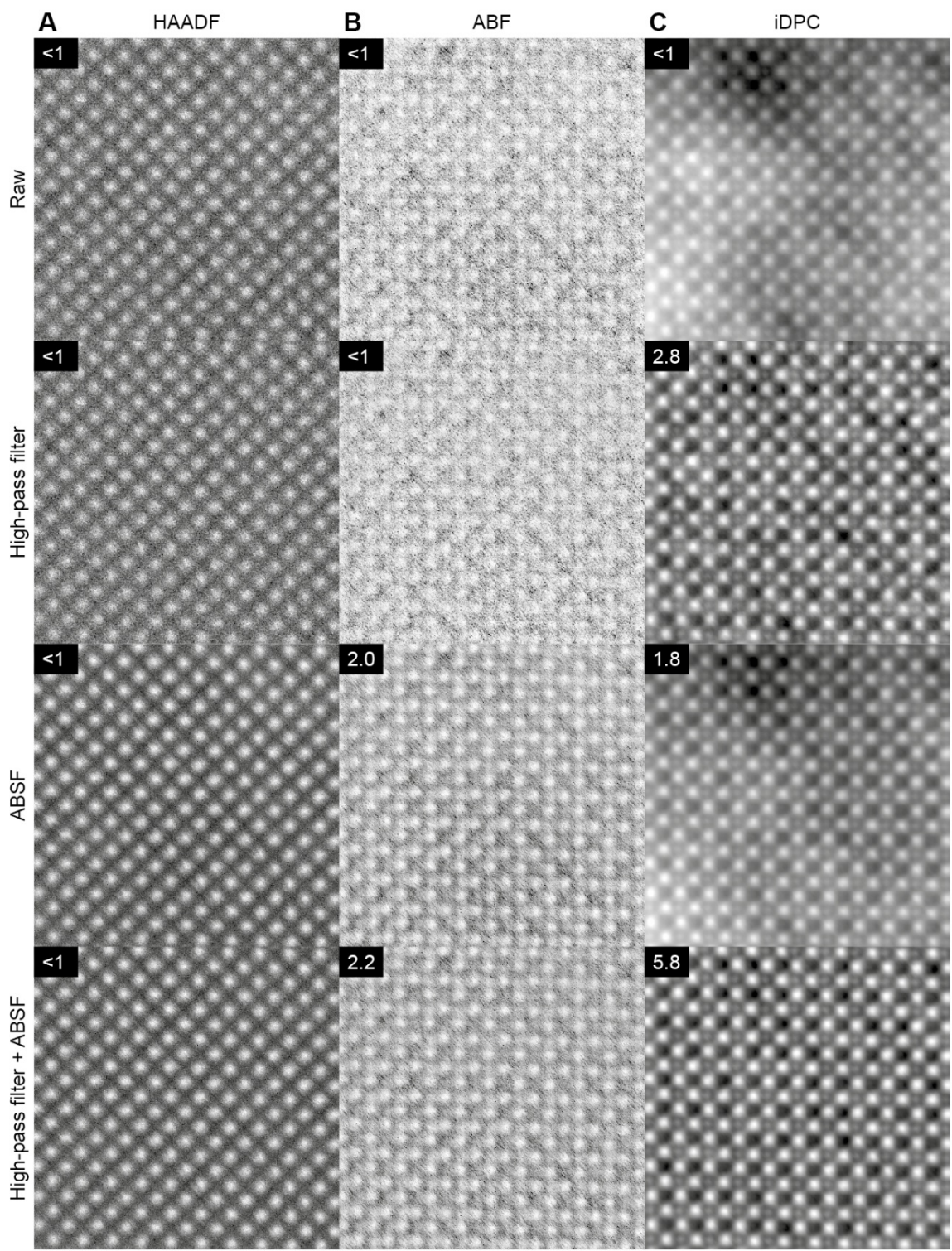

***Fig. S6***. ***Effect of the applied spatial filter on images of γ-TiH.*** *(**A**) HAADF (**B**) ABF (**C**) iDPC images. The average signal-to-noise ratio of the hydrogen signal is shown in the top left corner of each image.*

## 5. Reciprocal space analysis of the hydrogen signal in γ-TiH

We compare the Fourier Transforms (FT) of the images in Fig. 2A-D of the main article to investigate the strength of the signal from the hydrogen atoms. In **Fig. S7** we show the fast FT (FFT) of the full source images, where the symmetry of the α-Ti and γ-TiH lattices are also indicated. The checkerboard-like ordering of the hydrogen columns in γ-TiH is present as superlattice reflections in the FFTs of the images. The relative intensity of the γ-TiH{110} superlattice reflections with respect to a reflection from the Ti lattice (here we use the γ-TiH{200} peaks as reference) indicates how well the long-range ordering of hydrogen columns is defined in the image. The four intensity profiles that connect the γ-TiH{200} peaks are averaged and shown in Fig. S7E. From these peaks the relative {110}/{200} intensity is calculated and inset in the FFTs.

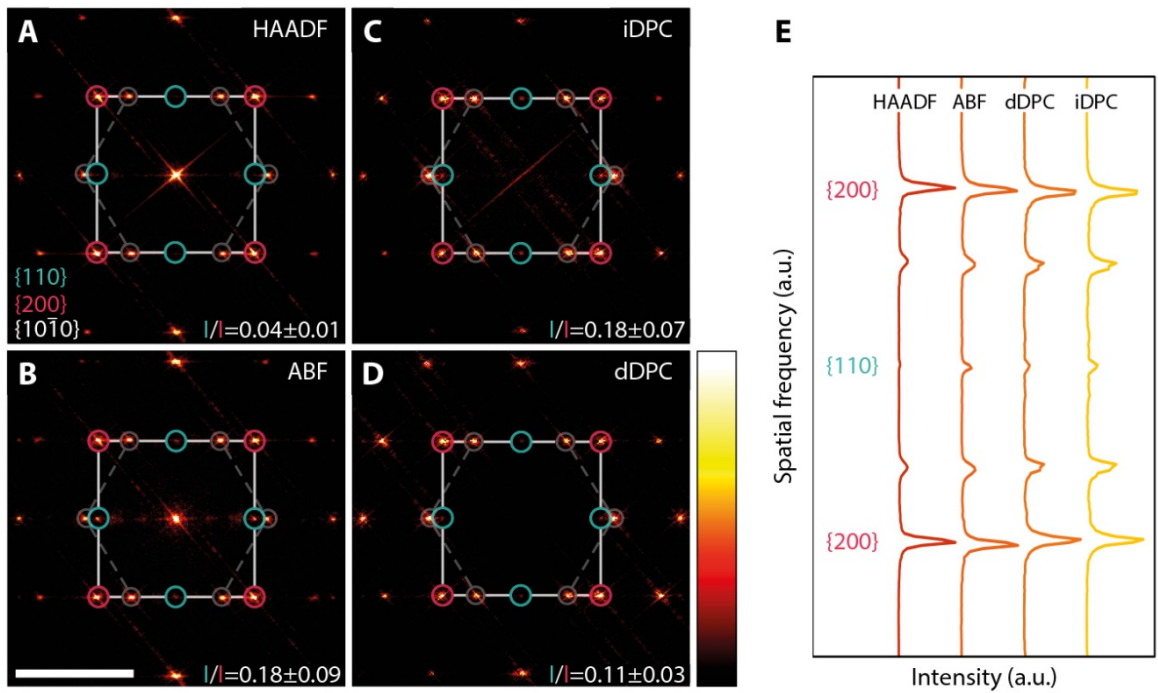

*Fig. S7. Reciprocal space analysis of the hydrogen signal. (A) HAADF, (B) ABF, (C) iDPC, (D) dDPC image Dashed and solid lines connect the encircled low-index reflections from the HCP α-Ti and FCT γ-TiH lattices, respectively. Scale bar size is 5.0 nm$^{-1}$. (E) Averaged intensity profiles of the four lines connecting the γ-TiH{200} reflections encircled in red, from which the relative intensity γ-TiH{110}/γ-TiH{200} is calculated and inset in the bottom-right of (A-D).*

In the case of the HAADF image there is no {110} peak present, but only a minor amount of streaking intensity that originates from the nearby α-Ti{10-10} peaks. In the dDPC image the relative intensity is close to half the value of the ABF and iDPC images. The ABF and iDPC images contain the highest relative strength of the super-reflections and are remarkably similar, indicating that the long-range {110} ordering is equally present in both images. Despite this similarity of the ABF and iDPC images in reciprocal space, the iDPC image has a significantly better short-range transfer of intensity of the hydrogen columns in real space (compare Figs. 2B and C of the main article).

The apparent discrepancy between the reciprocal space analysis and real-space analysis is caused by an imaging artefact in the ABF image. This is pointed out in the main article with Fig. 3B. There the average intensity profiles reveal that the Ti intensities are also modulated with a {110} periodicity. Hence, the combined effect of the hydrogen ordering in the γ-TiH crystal and the image artefact of the Ti atom column intensities, lead to the increased intensity of the {110} reflection in the FFT of the ABF image. Hence, the actual {110} intensity from the hydrogen atoms in the ABF image is lower than the iDPC image.

## 6. Multislice scanning transmission electron microscopy simulations

The three-dimensional models of γ-TiH and its interface with α-Ti are constructed with VESTA (*45*). The lattice parameters of freestanding γ-TiH are set to a=4.20 Å and c=4.60 Å and those of α-Ti to a=2.9508 Å and c=4.6855 Å (*29*). For the models of the interface the γ-TiH unit cell is slightly adjusted by -0.6% for the a lattice parameter (a=4.1730 Å) and +1.9% for the c lattice parameter (c=4.6855 Å) to obtain a perfect coherent interface with the α-Ti where the interface plane is formed by $(01\text{-}10)_\alpha//(1\text{-}10)_\gamma$ and where in-plane the $[0001]_\alpha//[001]_\gamma$ (and $[\text{-}2110]_\alpha//[110]_\gamma$) directions are parallel.

These models are loaded in software for multislice simulations called Dr. Probe (*46*). The models are always one unit cell thick and are cut in four equally thick slices. With this approach each slice contains one atomic plane in the case of γ-TiH (alternating Ti and H planes), whereas the α-Ti phase contains two empty slices and two Ti planes (alternating Ti and empty planes) as no atoms are present at z=1/4 and z=3/4. Atomic vibrations are accounted for with the frozen-lattice method, where the used Debye-Waller factors at 300 K are 0.52 Å$^2$ for Ti and 5.0 Å$^2$ for H. The factor for Ti is taken from the table in the work of Peng et al. (*47*), while we linearly extrapolated the Debye-Waller factor for H from the same table using the listed group I elements. It is assumed that the microscope has no aberrations except for defocus, and the operating conditions are set equal to the experimental settings of 300 kV and a convergence angle of 21 mrad. In experiment, the corrector software reduces the high order aberrations to values close to zero, the first order aberrations (A1 and C1) to values below 0.5 nm and second order aberrations (A2 and B2) to values in the range of 5 nm. The probe step size is set to 10 pm or lower, depending on the size of the simulated image. The used detectors are given in **Table S1**, which are equivalent to the calibrated experimental values. The final simulated ABF and HAADF images are then convolved with a two dimensional Gaussian function of 70 pm FWHM to account for the finite probe size. In addition to the normal ABF image, we also add the images from the four quadrant detectors to obtain an ABF-like image that we call DPC sum. The difference is that the inner collection angle is a bit smaller, and the outer collection angle extends into the dark field part of the diffraction pattern.

| Detector name | Azimuthal range (°) | Inner collection angle (mrad) | Outer collection angle (mrad) |
| --- | --- | --- | --- |
| DPC_A | -45 – 45 | 10 | 40 |
| DPC_B | 45 – 135 | 10 | 40 |
| DPC_C | 135 – 225 | 10 | 40 |
| DPC_D | 225 – 315 | 10 | 40 |
| ABF | 0 – 360 | 11 | 18 |
| HAADF | 0 – 360 | 43 | 200 |

**Table S1.** *List of the electron detectors and their corresponding collection angles that are used in experiment and simulation.*

The images from the DPC_(A-D) detectors, however, are first processed to obtain the iDPC image, according to the method described in the work of Lazic et al. (*20*). This processing procedure is as follows: first the opposite segments are subtracted (i.e. DPC_A – DPC_C and DPC_B – DPC_D, in the detector coordinates system) to obtain the orthogonal (x,y) differential phase contrast (DPC) images $DPC_x$ and $DPC_y$ by rotating the vector components to match the scanning (image) coordinates system (*37*). Periodic boundary conditions are constructed for the (non-periodic) interface by mirroring the $DPC_x$ and $DPC_y$ images, which is necessary to prevent aliasing effects for the simulations that cover a small field-of-view (*48*). Next, the $DPC_x$ and $DPC_y$ images are integrated in the Fourier domain to obtain the Fourier transformed iDPC image:

$$\mathcal{F}\{iDPC(x,y)\}(k_x,k_y) = \frac{k_x \cdot \mathcal{F}\{DPC_x(x,y)\}(k_x,k_y) + k_y \cdot \mathcal{F}\{DPC_y(x,y)\}(k_x,k_y)}{2\pi i(k_x^2 + k_y^2)}$$

Where $k_x$ and $k_y$ are the orthogonal k-vector components in the Fourier domain. The intensity in the Fourier transformed iDPC image at $k_x = k_y = 0$ is set to 0 to account for the singularity, before inverse Fourier transforming to obtain the real space iDPC image. Note that this is merely a constant that shifts the intensity of the image. Next, the iDPC image is convolved with the two dimensional Gaussian function of 70 pm FWHM to account for the finite probe size.

The iDPC image that is corrected for the finite probe size is used to obtain the DPC vector image and dDPC scalar image. The DPC image is obtained from the gradient of the iDPC image, which returns the two in-plane (x,y) vector components of the DPC image. This provides a physically regularized DPC image, where the non-conservative part of the noise is removed (*20, 37*). Differentiating once more, by applying the divergence operator to the DPC vector image, yields the dDPC scalar image.

# 7. Multislice simulations of the γ-TiH unit cell with and without hydrogen atoms

In this section we show that the observed peaks in the experimental images, as depicted in Figure 4 of the main article, are indeed a result of the hydrogen atoms, instead of being an image artifact that stems from interference effects. For this purpose, we have simulated the γ-TiH unit cell with hydrogen atoms and without hydrogen atoms. In **Figure S8** we present the simulated ABF, DPC sum and iDPC images for the empty (Figure S8A) and filled (Figure S8B) γ-TiH unit cell. In this figure the intensity profile along the face diagonal (Figure S8A), or a combined result of the two face diagonals (Figure S8B) is mapped as a function of thickness. In this way the intensity profile contains three Ti atomic columns and two empty columns in Figure S8A, and three Ti atomic columns, one empty column and one hydrogen atomic column in Figure S8B. The lower half of the figure shows the simulated images for a selection of specimen thickness' in steps of 9.2 nm (or 20 unit cells).

In case of the ABF and DPC sum images, four broad peaks are present between the Ti columns in Figure S8A for crystals thinner than about 12 nm and 6 nm, respectively. The iDPC image also shows this feature but for crystals with a thickness around 15-25 nm. This demonstrates that when no hydrogen atoms are present, the erroneous conclusions can be drawn that a dihydride crystal has been imaged, where the four tetrahedral columns are filled with hydrogen atoms.

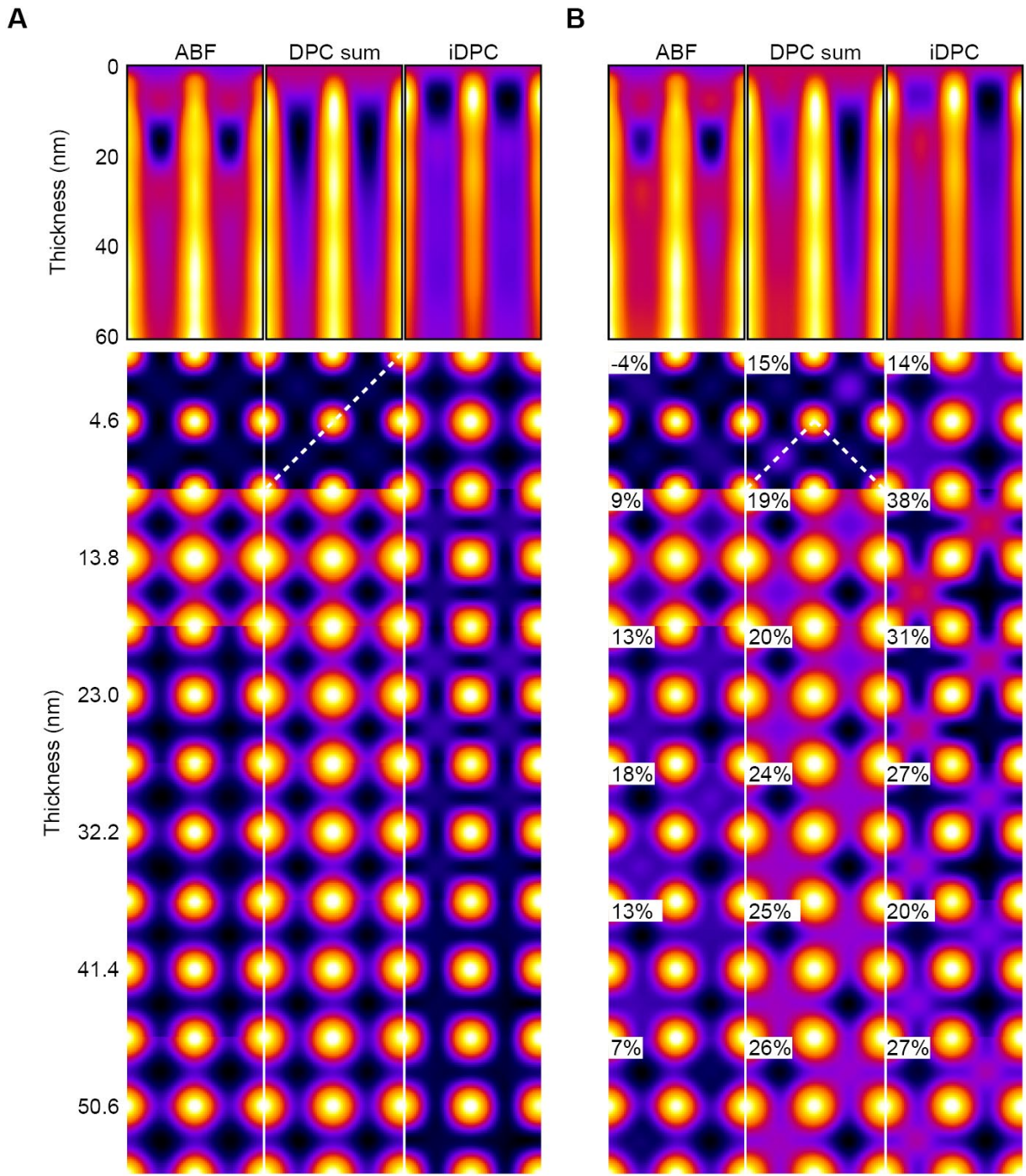

***Fig. S8***. ***Image simulations of the γ-TiH unit cell with and without hydrogen atoms.*** *Simulated ABF (Δf=0 nm), DPC sum (Δf=0 nm) and iDPC (Δf=-4 nm) images of the γ-TiH unit cell images are compared. At the top part the intensity profiles (along the indicated white dashed line indicated below) as a function of thickness are shown. The bottom part shows the γ-TiH unit cell (**A**) without hydrogen atoms, and (**B**) with hydrogen atoms for various thicknesses. In (**B**) the relative hydrogen intensity is inset. The contrasts of the ABF and DPC sum images are inverted to ease the visual comparison.*

The simulated images of the normal γ-TiH unit cell (Figure S8B) show that the hydrogen columns can be imaged by the three methods since relatively bright peaks at the expected position are observed in all cases. The highest relative intensity of the hydrogen columns is achieved with iDPC, followed by DPC sum and least with ABF. In the case of ABF the crystal should be thinner than about 35 nm, otherwise the intensity from the hydrogen column is drastically reduced. The DPC sum image is able to transfer the signal from the hydrogen column with a rather stable contrast even for relatively thick specimen. However, the intensity at the location of the hydrogen column does not resemble a well-defined peak. This is clearly better in case of iDPC, where the contrast of the hydrogen column is the highest, and also a well-defined peak shape occurs.

Results like shown Figure S8B hold for single defocus values. We repeated this for different defocus values and quantitatively analyzed and plotted all results in **Figure S9**, **Figure S10** and **Figure S11**, for ABF, DPC sum and iDPC imaging, respectively. In Figure S9A,B the relative intensity of the hydrogen column and titanium column w.r.t. the empty column is plotted. This indeed shows that for ABF imaging the contrast reversal of the hydrogen and empty column at zero defocus (which is the operating value in experiment) for crystals thinner than about 10 nm. For more negative defocus, the contrast reversal increases in magnitude and extends to slightly thicker crystals, and also the contrast between the titanium column and empty column is reversed. For more positive defocus the contrast between the titanium column and empty column is again reversed. Hence, for ABF imaging it is crucial to stay close (Δf=±4 nm or smaller) to zero defocus.

The same result is shown in Figure S10, but now for the DPC sum image. This result is similar to the case of ABF, but with slightly less oscillatory behavior. The contrast reverses between the hydrogen and empty column only for negative defocus values for crystals thinner than about 10-20 nm. However, the contrast reverses between the titanium column and the empty column for both positive and negative defocus values. In Figure S10C it is also apparent that the relative intensity of the hydrogen column w.r.t. the Ti column is remarkably independent from the crystal thickness, as has been shown in Figure S8B. However, also for DPC sum imaging it is crucial to stay close (Δf=±4 nm or smaller) to zero defocus.

Finally, the analysis of the iDPC imaging is shown in Figure S11. Contrast reversals between the titanium and empty column occur for all positive defocus values, and occur between the hydrogen and empty column for positive defocus values larger than 4 nm. However, for negative defocus values there are no contrast reversals, except for crystals thinner than about 4 nm for the most negative defocus value of -8 nm that we simulated. Physically this means that in the case of iDPC it is important to focus the probe into the specimen i.e. below the top surface. The value of the defocus is only limited by the specimen thickness, as it is the depth at which the projected electrostatic potential is probed by the converged beam waist allowing atomically resolved 3D iDPC imaging (*49*). In addition, it shows enhanced contrast between the hydrogen and titanium columns as plotted in Figure S11C.

The above image simulations clearly demonstrate that iDPC imaging provides better results than ABF and DPC sum imaging with best localization and highest contrast of the hydrogen column, largest thickness and defocus ranges without disturbing contrast reversals and oscillatory behavior.

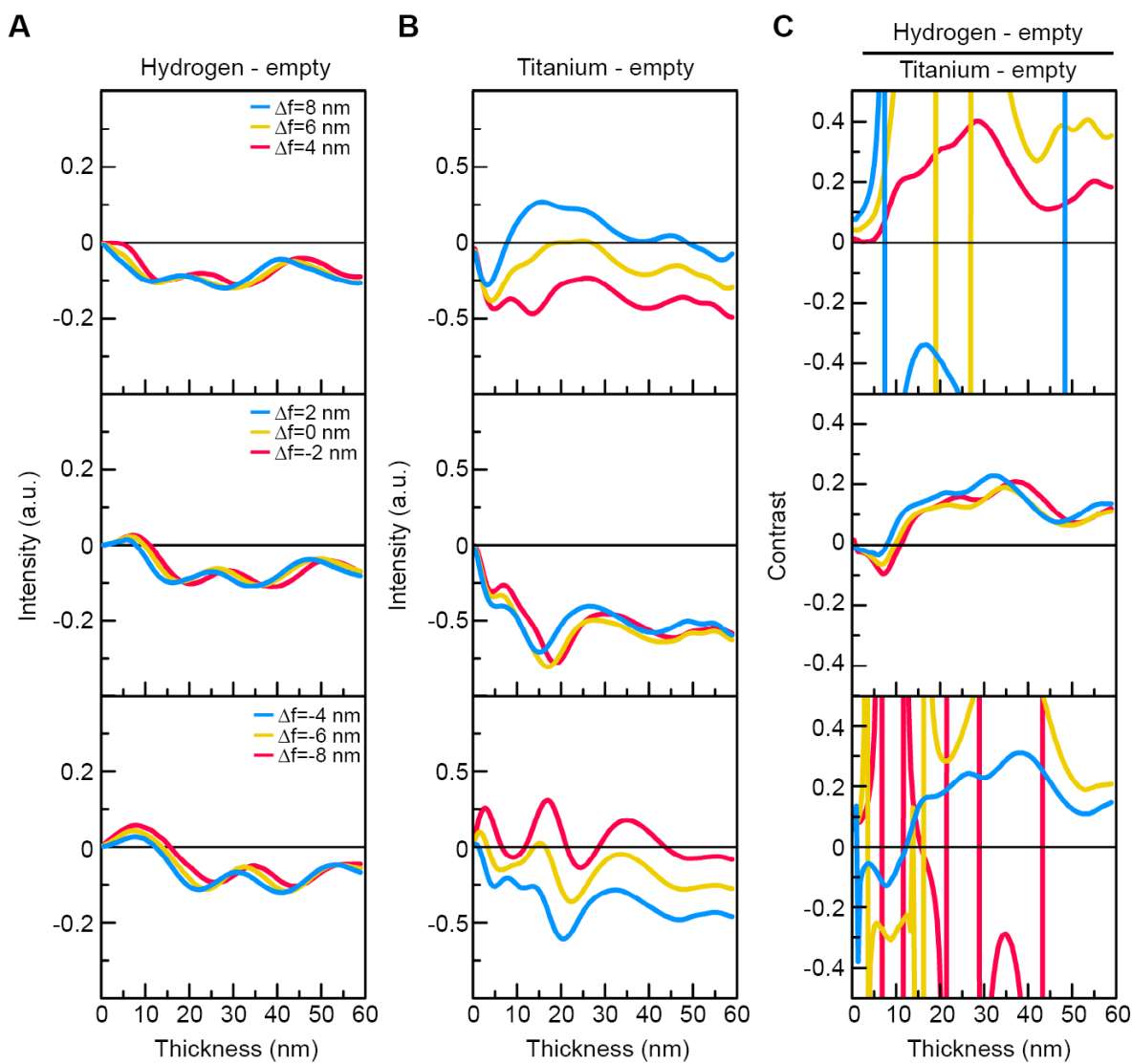

*Fig. S9. Simulated relative intensity of the hydrogen and titanium atoms in the ABF image.* *The relative intensities of the (A) hydrogen and (B) titanium column w.r.t. the empty column are plotted as a function of crystal thickness for various defocus settings. The ratio of the former two defines the relative intensity of the hydrogen column in (C). We classified the results in three defocus windows: negative (-8 nm to -4 nm), close to zero (-2 nm to +2 nm), positive (+4 nm to +8 nm). The vertical scale limits are (A) -0.4 to 0.4 (B) -1.0 to 1.0 (C) -0.5 to 0.5 and allow direct comparison of Figures. S9, S10 and S11.*

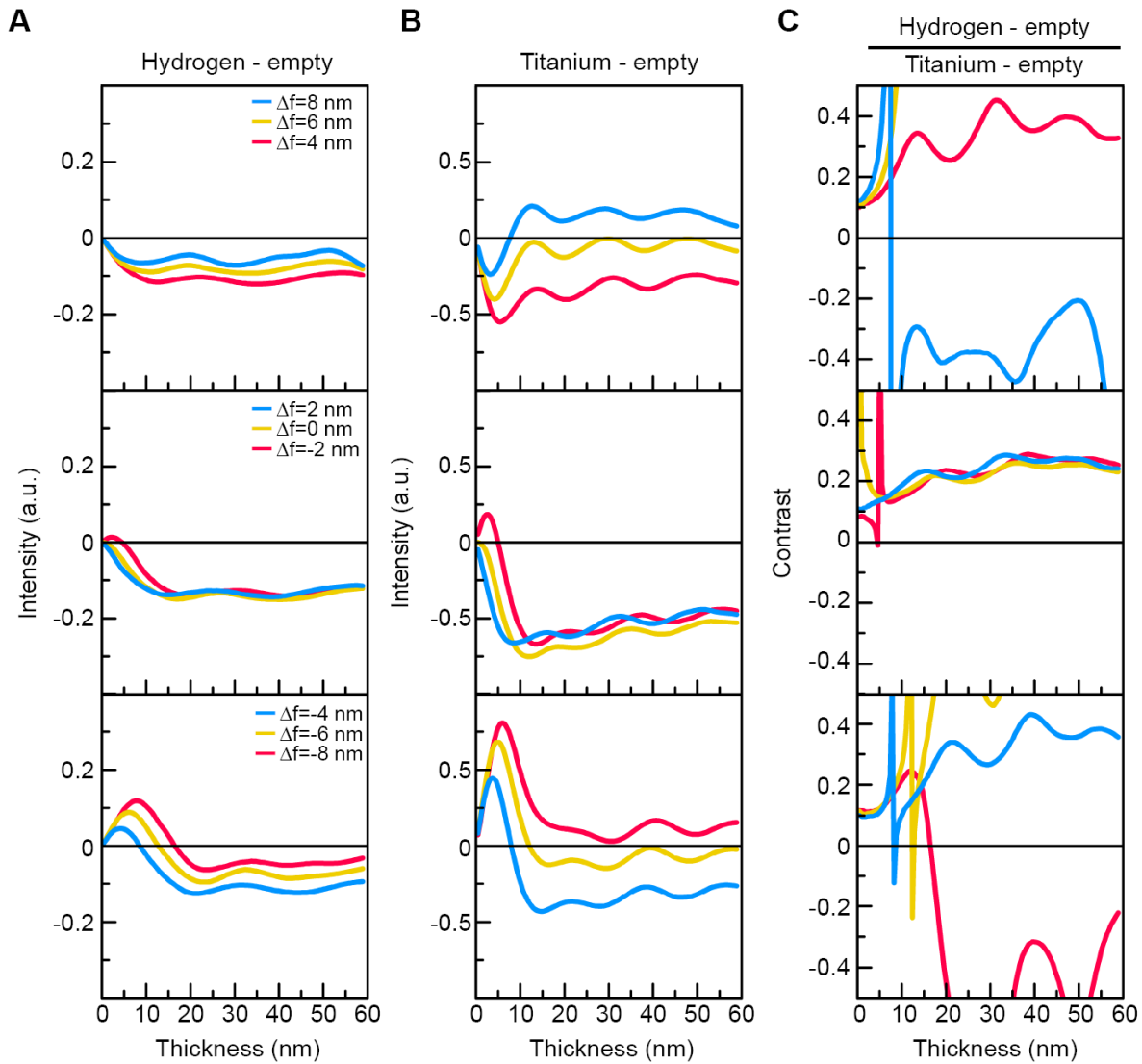

***Fig. S10**. **Simulated relative intensity of the hydrogen and titanium atoms in the DPC sum (ABF-like) image.*** *The relative intensities of the (**A**) hydrogen and (**B**) titanium column w.r.t. the empty column are plotted as a function of crystal thickness for various defocus settings. The ratio of the former two defines the relative intensity of the hydrogen column in (**C**). We classified the results in three defocus windows: negative (-8 nm to -4 nm), close to zero (-2 nm to +2 nm), positive (+4 nm to +8 nm). The vertical scale limits are (**A**) -0.4 to 0.4 (**B**) -1.0 to 1.0 (**C**) -0.5 to 0.5 and allow direct comparison of Figures. S9, S10 and S11.*

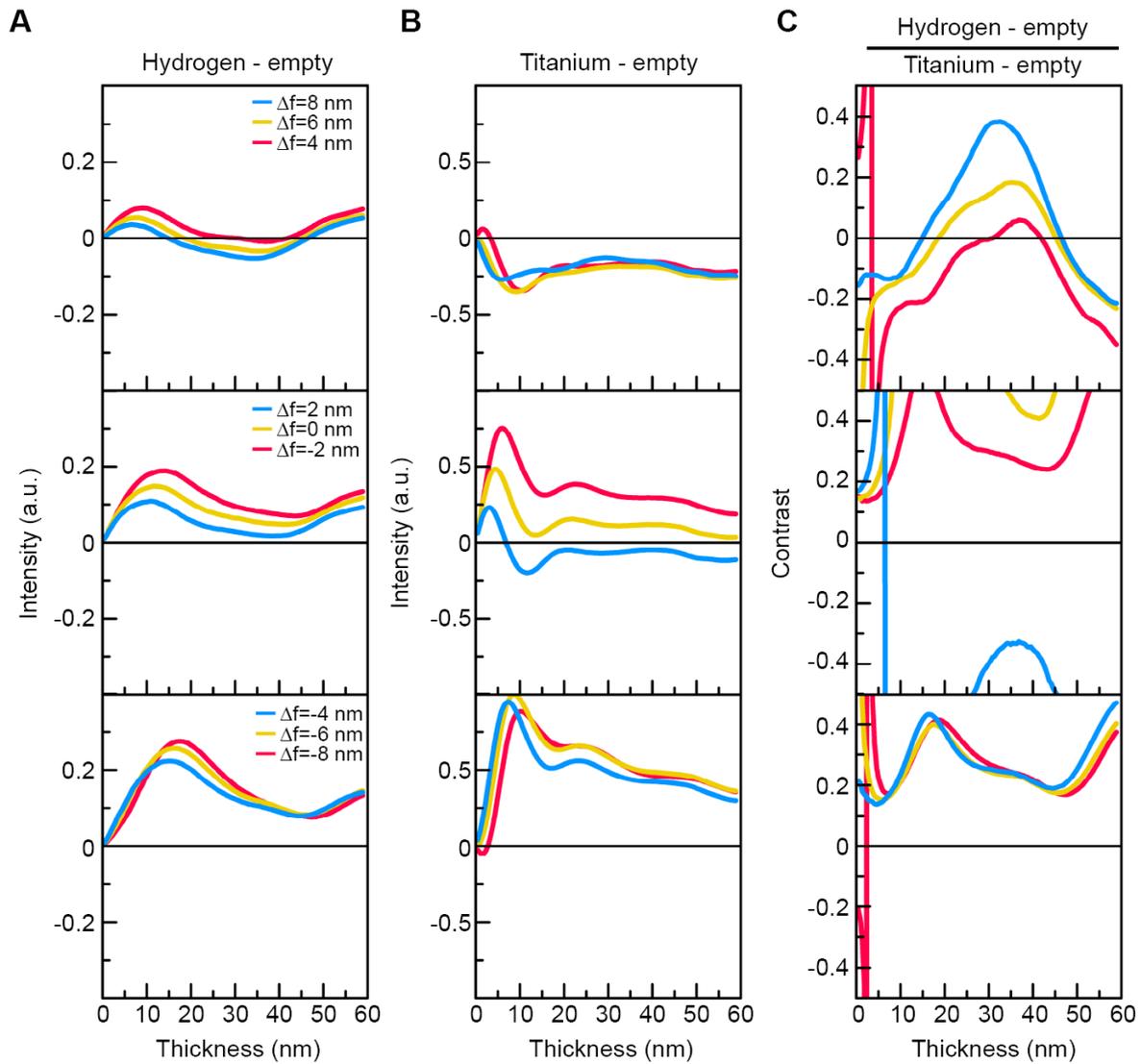

***Fig. S11***. ***Simulated relative intensity of the hydrogen and titanium atoms in the iDPC image.*** *The relative intensities of the (**A**) hydrogen and (**B**) titanium column w.r.t. the empty column are plotted as a function of crystal thickness for various defocus settings. The ratio of the former two defines the relative intensity of the hydrogen column in (**C**). We classified the results in three defocus windows: negative (-8 nm to -4 nm), close to zero (-2 nm to +2 nm), positive (+4 nm to +8 nm). The vertical scale limits are (**A**) -0.4 to 0.4 (**B**) -1.0 to 1.0 (**C**) -0.5 to 0.5 and allow direct comparison of Figures. S9, S10 and S11.*

## 8. Multislice simulations of the tilted γ-TiH unit cell with and without hydrogen atoms

In the experiments the local alignment of the specimen was constrained by the small length scale crystal bending. From the experimental images usually only a small part was well aligned, which we assessed based on the combined quality of the HAADF and iDPC or ABF images. The combined effect of crystal tilt and the wave character behavior of particularly the iDPC and ABF imaging could potentially induce imaging artefacts that lead to erroneous detection of atomic columns. Therefore, we investigate if crystal tilt can introduce peaks in the image that could be wrongly attributed to hydrogen atomic columns.

To this purpose we have performed multislice image simulations of a tilted γ-TiH unit cell, with and without the hydrogen atoms. We applied a crystal tilt of 0.5° (8.7 mrad) in x and y direction (i.e. the unit cell is tilted along a face diagonal), as these were approximately the maximum misalignments we encountered in experiments, especially near the interface. The reason why we chose to tilt the crystal along the face diagonal is that crystal tilt in the x or y direction would certainly not generate the asymmetry of intensity peaks that we observe i.e. this will not produce the ordered {110} type of intensity peaks.

We present the results of these image simulations in **Figure S12** in the same way as we presented Figure S8. However, now we show the intensity profile of the entire face diagonal only in the tilt direction in Figure S12B, instead of the combined profile of both diagonals like we presented in Figure S8B. We cannot use the intensity profile of the other (anti-)diagonal due to streaking of the titanium atomic columns. The intensity maps at the top panel can be used to compare the empty (Figure S12A) and hydrogen (Figure S12B) columns.

For the three imaging techniques the titanium atomic columns in Figure S12A have asymmetric shapes as a consequence of the crystal tilt. The titanium atomic columns also appear to be displaced with respect to their actual position. This effect is mostly visible for ABF and DPC sum, and only minimally for iDPC where the atomic columns remain most localized. Furthermore, the apparent position of the atomic columns is very sensitive to thickness and can show relatively large jumps for minor increases in crystal thickness. This is most visible for DPC sum and ABF for about 15 nm thick crystal, and again at around 40 nm.

More relevant is that the empty unit cell (see Figure 12A) contains apparent atomic columns between the titanium columns only for the DPC sum image, and not for ABF and iDPC images. When the crystal is not tilted, there are apparent atomic columns for all three imaging techniques. Hence, when the crystal is tilted, interference effects are reduced for ABF and iDPC, but not for DPC sum. When hydrogen atoms are present (see Figure S12B) then only DPC sum and iDPC show visible contrast between the hydrogen and empty column, and ABF does not. The resulting checkerboard-like pattern remains most clearly visible when the crystal is thinner than about 30 nm for DPC sum imaging, as the contrast reduces for thicker crystals. Whereas the contrast remains best for iDPC imaging even for thick crystals.

Overall, these results demonstrate that iDPC imaging is least sensitive towards crystal tilt on several aspects. iDPC imaging shows the least atomic column displacement, and remains capable of imaging hydrogen atomic columns even for thick crystals despite the rather extreme crystal tilt. In contrast, ABF and DPC sum imaging show substantial displacement and also sudden jumps in apparent atomic column position. Furthermore, ABF imaging is not able to transfer any significant signal from the hydrogen columns and DPC sum imaging can do this only for crystals thinner than about 30 nm.

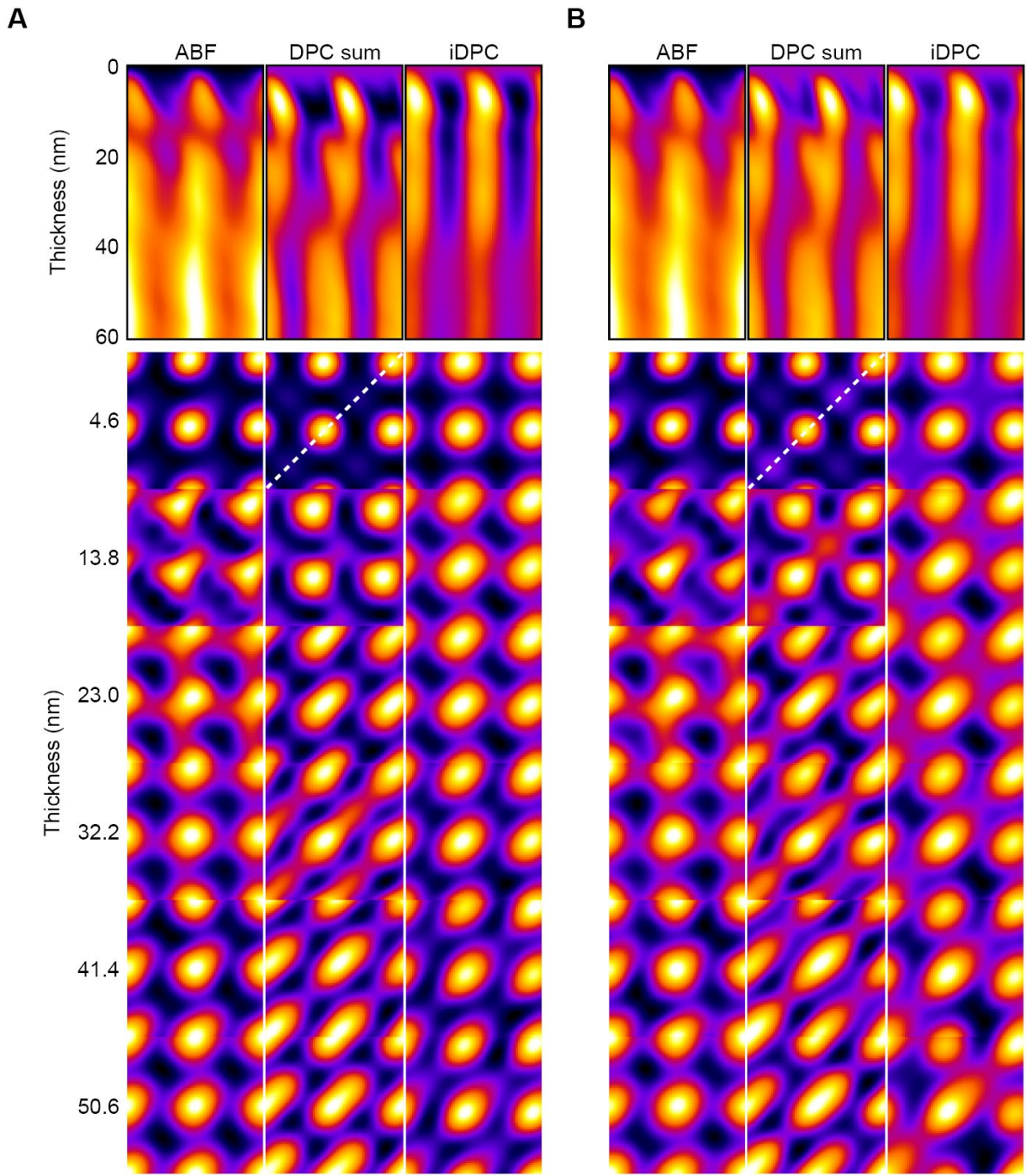

***Fig. S12**. **Image simulations of the tilted γ-TiH unit cell with and without hydrogen atoms.*** Simulated ABF (*Δf=0 nm*), DPC sum (*Δf=0 nm*) and iDPC (*Δf=-4 nm*) images of the tilted γ-TiH unit cell images are compared. On the top the intensity profiles (along the indicated dashed line below) as a function of thickness are shown. The bottom shows the γ-TiH unit cell (***A***) without hydrogen atoms and (***B***) with hydrogen atoms. The contrast of the ABF and DPC sum images are inverted to ease the visual comparison.

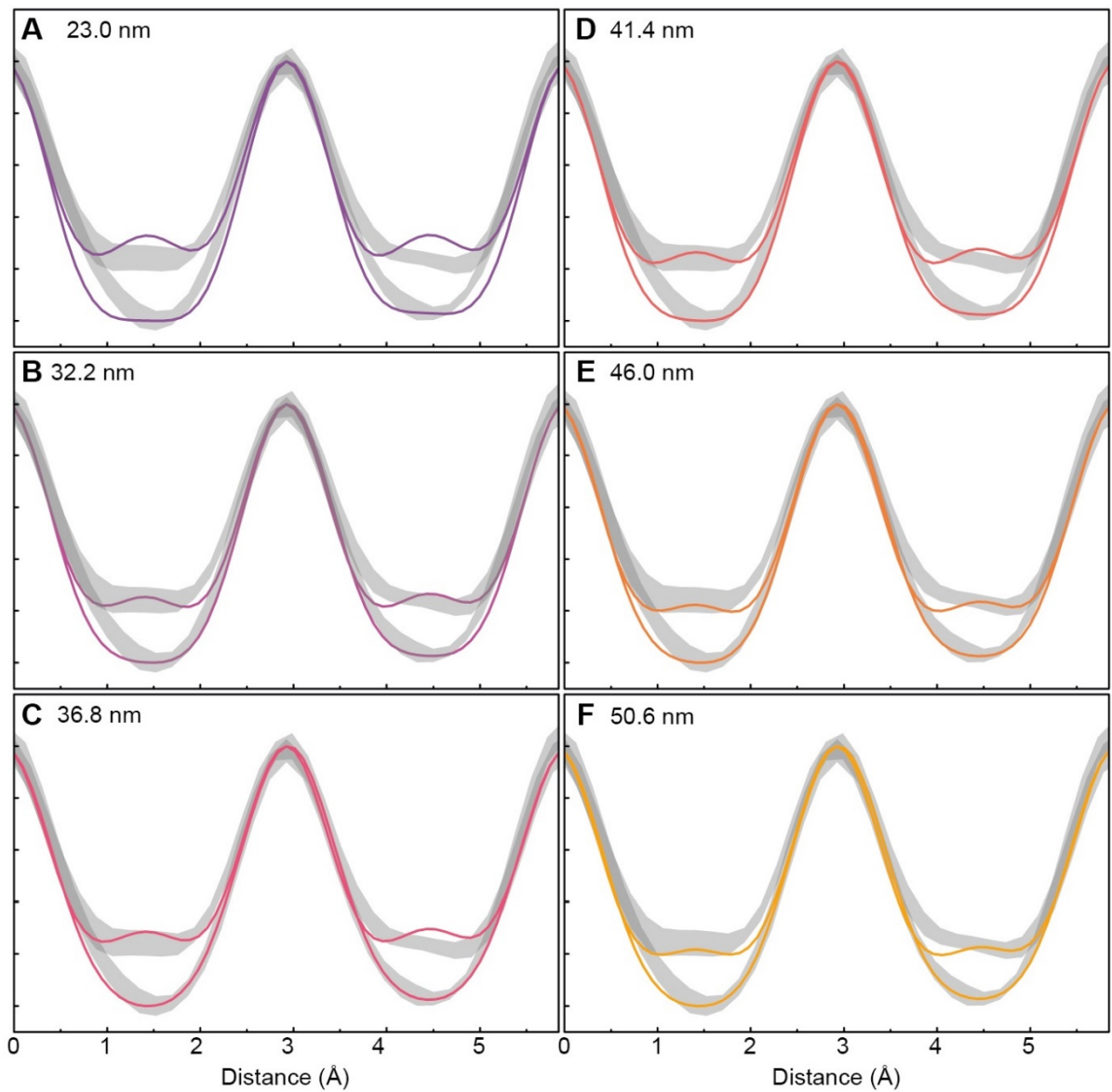

*Fig. S13. Comparison between experimental and simulated intensity profiles of the iDPC image of γ-TiH at the interface with Ti.* Normalized intensity profiles from the simulated (solid colored lines) iDPC image ($\Delta f$=-4 nm) are compared with the line profiles from the experimental (gray bands) image of Fig. 2c from the main article. The simulated specimen thicknesses are *(A)* 23.0 nm *(B)* 32.2 nm *(C)* 36.8 nm *(D)* 41.4 nm *(E)* 46.0 nm and *(F)* 50.6 nm. Vertical scale ranges from -0.1 to 1.2.